\documentclass[12pt]{article}
\usepackage{graphicx}
\usepackage{amsmath}
\usepackage{amssymb}
\usepackage{caption2}
\begin{document}
\bibliographystyle{prsty}
\begin{center}
{\large {\bf \sc{ Analysis of the $\Sigma-n $  form-factors
with light-cone QCD sum rules  }}} \\[2mm]
Z. G. Wang \footnote{E-mail,wangzgyiti@yahoo.com.cn.  }     \\
 Department of Physics, North China Electric Power University, Baoding 071003, P. R. China \\
\end{center}

\begin{abstract}
In this work, we  study  the four form-factors
   $f_1(Q^2)$, $f_2(Q^2)$, $g_1(Q^2)$ and $g_2(Q^2)$ of the $\Sigma \to n$ in the
  framework of the light-cone QCD sum rules approach up
 to twist-6 three valence quark light-cone distribution amplitudes.
 The $f_1(0)$ is the basic
input parameter in extracting the CKM matrix element $|V_{us}|$ from
the hyperon decays.
  The  four form-factors $f_1(Q^2)$, $f_2(Q^2)$, $g_1(Q^2)$ and $g_2(Q^2)$ at intermediate and
large momentum transfers with $Q^2> 3 GeV^2$ have significant
contributions from the end-point (soft) terms.  The numerical values
of  the four form-factors
   $f_1(0)$, $f_2(0)$, $g_1(0)$ and $g_2(0)$  are
 compatible with the experimental data and  theoretical
 calculations (in magnitude); although the uncertainties are large.
\end{abstract}

PACS numbers:  12.38.Lg; 12.38.Bx; 12.15.Hh

{\bf{Key Words:}}  Form-factors, CKM matrix element, light-cone QCD
sum rules
\section{Introduction}
Semileptonic decays $K\to\pi\ell \nu$ ($K_{\ell 3}$)  provide the
most precise determination of the Cabibbo-Kobayashi-Maskawa (CKM)
matrix element $|V_{us}|$ \cite{CKM}. The experimental input
parameters are the semileptonic decay widths and the vector
form-factors $f^+_{K\pi}(q^2)$ and $f^-_{K\pi}(q^2)$, which  are
necessary in calculating the phase space integrals. The main
uncertainty in the quantity $|V_{us}f^+_{K\pi}(0)|$ comes from the
unknown shape of the hadronic form-factor $f^+_{K\pi}(q^2)$, which
is measurable at $m_l^2<q^2<(m_K-m_\pi)^2$ in the $K_{\ell 3}$
 decays or $(m_K+m_\pi)^2<q^2<m^2_\tau$  in the $\tau \to K\pi\nu$
decays. Another way to extract the $|V_{us}|$ is provided by the
hyperon semileptonic decays,  it is possible to extract the quantity
$|V_{us} f_1(0)|$ at the percent level from the hyperon experiments
\cite{Cabibbo03}, where the $f_1(0)$ is the vector form-factor at
zero-momentum transfer. The Ademollo-Gatto   theorem protects the
$f_1(0)$ from the first-order $SU(3)$-breaking corrections
\cite{Gatto64}, while the second-order corrections   are badly
known. There exist several model dependent estimates for the
$f_1(0)$, for examples,  quark models \cite{quarkmodel}, large-$N_c$
\cite{largeN} and chiral expansions\cite{Chiral}; however, the
values disagree with each other. The axial-vector form-factor
$g_1(0)$ is not protected by the Ademollo-Gatto theorem, and it
suffers from the first-order $SU(3)$ breaking corrections. The $
SU(3)$ symmetry  implies a vanishing "weak-electricity" form-factor
$g_2(Q^2)$, because charge conjugation does not allow a  $C$-odd
 term $g_2(Q^2)$ in the matrix elements of the neutral axial-vector
currents $A_\mu^3$ and $A_\mu^8$, which are $C$-even.

In our previously work, we study the vector form-factors
$f^+_{K\pi}(q^2)$ and $f^-_{K\pi}(q^2)$ with the light-cone QCD sum
rules (LCSR),  and obtain satisfactory results \cite{WangWan06}.  In
this article, we calculate the   four form-factors
   $f_1(Q^2)$, $f_2(Q^2)$, $g_1(Q^2)$ and $g_2(Q^2)$ of the $\Sigma \to n$  in the
framework of the LCSR approach \cite{LCSR, LCSRreview}, which
combine the standard techniques  of the QCD sum rules with the
conventional parton distribution amplitudes describing  the hard
exclusive processes\cite{SVZ79}. In the LCSR approach, the
short-distance operator product expansion with   the vacuum
condensates of increasing dimensions is replaced by the light-cone
expansion with the distribution amplitudes (which correspond to the
sum of an infinite series of operators with the same twist) of
increasing twists  to parameterize the non-perturbative  QCD vacuum.
The higher twists light-cone distribution amplitudes   of the
baryons were not available until recently \cite{BFMS}, then the
LCSRs were applied to study the form-factors of the nucleons
\cite{BaryonBraun,WangWY,Braun06} and the weak decays
\cite{Huang04}.

The article is arranged as: in Section 2, we derive the analytical
expressions of the four form-factors  $f_1(Q^2)$, $f_2(Q^2)$,
$g_1(Q^2)$ and $g_2(Q^2)$ with the light-cone QCD sum rules
approach; in Section 3, the numerical results and discussions; and
in Section 4, conclusion.

\section{Form-factors  $f_1(Q^2)$, $f_2(Q^2)$, $g_1(Q^2)$ and $g_2(Q^2)$   with light-cone QCD sum rules}

In the following, we write down  the two-point correlation function
$\Pi_\mu(P,q)$ in the framework of the LCSR approach,
\begin{eqnarray}
 \Pi_\mu(P,q) &=& i \int d^4 x \, e^{-i q \cdot x}
\langle 0| T\left\{\eta(0) J_\mu (x)\right\} |P\rangle ,
\end{eqnarray}
 with the chiral current
\begin{eqnarray}
 J_\mu(x)&=&  \bar{s}(x)\gamma_\mu(1-\gamma_5)  u(x)  ,
\end{eqnarray}
 and the Ioffe  current for the $\Sigma$ baryon
\cite{Ioffe}
\begin{eqnarray}
 \eta(0) &=& \epsilon^{abc} d^T_a(0) C \gamma_\mu d_b(0) \, \gamma_5  \gamma^ \mu s_c(0) \,,
\nonumber\\
 \langle0| \eta(0)  |P\rangle & = & \lambda_{\Sigma}N(P)\, ,
\end{eqnarray}
 here  the  $\lambda_\Sigma$ is the
 coupling constant of
  the $\Sigma$ baryon.
There are two independent interpolating currents with the
spin-$\frac{1}{2}$ and isospin-1, both are expected to excite the
ground state $\Sigma$ baryon from the vacuum, the general form of
the $\Sigma$ current can be written as \cite{Baryon}
\begin{eqnarray}
\eta(x,t) &=&  \epsilon^{abc} \left\{ d_a^T(x) C  s_b(x)
\gamma_5d_c(x) + t  d_a^T(x) C \gamma_5 s_b(x)  d_c(x) \right \},
\nonumber
\end{eqnarray}
in the limit $t=-1$,   the Ioffe current is recovered,  we can take
the $t$ as a free parameter  and select the ideal value with the QCD
sum rules approach, here we prefer the Ioffe current $\eta(x)$
 to keep in consistent with the QCD sum rules used in
determining the parameters in the light-cone distribution
amplitudes. We can also choose the Chernyak-Zhitnitsky type current
to interpolate the $\Sigma$ baryon \cite{Che84}
\begin{eqnarray}
 \eta(x) &=& \epsilon^{abc} d^T_a(x) C \!\not\!{z} d_b(x) \, \gamma_5  \!\not\!{z} s_c(x)
 \,, \nonumber
\end{eqnarray}
where the $z_\mu$ is a light-cone four-vector, the currents of this
type have non-vanishing couplings to both the spin-$\frac{1}{2}$ and
-$\frac{3}{2}$ baryon states, it is difficult to separate the
contribution of the spin-$\frac{3}{2}$ state.

  At the large Euclidean momenta ${P'}^2=(P+q)^2$
and $q^2 = - Q^2$, the correlation function $\Pi_\mu(P,q) $ can be
calculated in perturbation theory. In calculation, we need the
following light-cone expanded quark propagator \cite{BB89},
\begin{eqnarray}
S(x) &=& \frac{i \not\!x}{2\pi^2x^4} -\frac{m}{4 \pi^2 x^2}\nonumber
\\
&-& \frac{i }{16\pi^2 x^2 }\int\limits_0^1dv \Big\{(1-v)\not\!x
\sigma_{\mu \nu} G^{\mu\nu}(vx) + v\sigma_{\mu \nu}
G^{\mu\nu}(vx)\!\not\! x \Big\} + ...,
 \end{eqnarray} where $
G_{\mu\nu}= g_sG_{\mu\nu}^a(\lambda^a/2)$ is the gluon field
strength tensor.  The contributions proportional to the $G_{\mu\nu}$
can give rise to four-particle (and five-particle) nucleon
distribution amplitudes with a gluon (or quark-antiquark pair) in
addition to the three valence quarks, their corrections are usually
not expected to play any significant roles \cite{DFJK} and neglected
here \cite{BaryonBraun,WangWY, Huang04}. In the parton model, at
large momentum transfers, the electromagnetic and weak currents
interact with the almost free partons in the nucleons. Employ the
"free" light-cone quark propagator in the correlation function
$\Pi_\mu(P,q)$, we obtain
\begin{eqnarray}
 &&z^\mu \Pi_\mu(P,q)\nonumber \\
 &=&   \int d^4 x \, \frac{ e^{-i q \cdot x}}{2 \pi^2 x^4}
(C\gamma_\mu)^{\alpha\beta} \left[\gamma^\mu
\!\not\!{x}\!\not\!{z}(1-\gamma_5)\right]^{\eta \lambda}
\epsilon_{ijk}\langle 0| T\left\{ d^i_\alpha(0)
d^j_\beta(0)u^k_\lambda(x)\right\} |P\rangle \nonumber \\
&+&  m_s i \int d^4 x \, \frac{ e^{-i q \cdot x}}{4 \pi^2 x^2}
(C\gamma_\mu)^{\alpha\beta}
\left[\gamma^\mu\!\not\!{z}(1-\gamma_5)\right]^{\eta \lambda}
\epsilon_{ijk}\langle 0| T\left\{ d^i_\alpha(0)
d^j_\beta(0)u^k_\lambda(x)\right\} |P\rangle. \nonumber
\end{eqnarray}

In the light-cone limit $x^2\to 0$, the remaining three-quark
operator sandwiched between the neutron state $|P\rangle$ and the
vacuum can be written in terms of the  nucleon distribution
amplitudes \cite{BFMS,Che84,earlybaryon}.  The three valence quark
components of the nucleon distribution amplitudes are defined by the
matrix element,
\begin{eqnarray}
&&4\langle0|\epsilon_{ijk}d_\alpha^i(a_1 x)d_\beta^j(a_2
x)u_\gamma^k(a_3 x)|P\rangle=
(\mathcal{V}_1+\frac{x^2M_n^2}{4}\mathcal{V}_1^M)(\rlap/PC)_{\alpha\beta}(\gamma_5N)_{\gamma}
\nonumber\\&&{}+
\mathcal{V}_2M_n(\rlap/PC)_{\alpha\beta}(\rlap/x\gamma_5N)_{\gamma}+
\mathcal{V}_3M_n(\gamma_\mu
C)_{\alpha\beta}(\gamma^\mu\gamma_5N)_{\gamma}+
\mathcal{V}_4M_n^2(\rlap/xC)_{\alpha\beta}(\gamma_5N)_{\gamma}\nonumber\\&&{}+
\mathcal{V}_5M_n^2(\gamma_\mu
C)_{\alpha\beta}(i\sigma^{\mu\nu}x_\nu\gamma_5N)_{\gamma} +
\mathcal{V}_6M_n^3(\rlap/xC)_{\alpha\beta}(\rlap/x\gamma_5N)_{\gamma}+\cdots.
\end{eqnarray}
The calligraphic distribution amplitudes do not have definite twist
and can be related to the ones with definite twist as
\begin{eqnarray}
&&\mathcal{V}_1=V_1, \hspace{2.4cm}2P\cdot
x\mathcal{V}_2=V_1-V_2-V_3, \nonumber\\&& 2\mathcal{V}_3=V_3,
\hspace{2.2cm} 4P\cdot
x\mathcal{V}_4=-2V_1+V_3+V_4+2V_5,\nonumber\\&& 4P\cdot
x\mathcal{V}_5=V_4-V_3,\hspace{0.5cm} (2P\cdot
x)^2\mathcal{V}_6=-V_1+V_2+V_3+V_4+V_5-V_6 \nonumber
\end{eqnarray}
for the vector distribution amplitudes. The light-cone distribution
amplitudes $F=V_i$ can be represented as
\begin{equation}
F(a_ip\cdot x)=\int dx_1dx_2dx_3\delta(x_1+x_2+x_3-1) e^{-ip\cdot
x\Sigma_ix_ia_i}F(x_i)\,  .
\end{equation}

The  distribution amplitudes are scale dependent and can be expanded
with the operators of increasing conformal spin, we write down the
explicit expressions for the $V_i$  up to the next-to-leading
conformal spin accuracy in the appendix \cite{BFMS}. The $V_1$ is
the leading twist-3 distribution amplitude; the  $V_2$ and $V_3$ are
the twist-4 distribution amplitudes; the   $V_4$ and $V_5$  are the
twist-5 distribution amplitudes; while the twist-6 distribution
amplitude is the $V_6$. The parameters $\phi_3^0$, $\phi_6^0$,
$\phi_4^0$, $\phi_5^0$, $\xi_4^0$, $\xi_5^0$, $\psi_4^0$,
$\psi_5^0$, $\phi_3^-$, $\phi_3^+$, $\phi_4^-$, $\phi_4^+$,
$\psi_4^-$, $\psi_4^+$, $\xi_4^-$, $\xi_4^+$, $\phi_5^-$,
$\phi_5^+$, $\psi_5^-$, $\psi_5^+$, $\xi_5^-$, $\xi_5^+$,
$\phi_6^-$, $\phi_6^+ $ in the light-cone distribution amplitudes
$V_i$ can be expressed in terms of eight independent matrix elements
of the local operators with the parameters $f_N$, $\lambda_1$,
$\lambda_2$, $V_1^d$, $A_1^u$, $f^d_1$, $f^d_2$ and $f^u_1$, the
three parameters $f_N$, $\lambda_1$ and $\lambda_2$ are related to
the leading order (or $S$-wave) contributions of the conformal spin
expansion,  the remaining five parameters $V_1^d$, $A_1^u$, $f^d_1$,
$f^d_2$ and $f^u_1$ are related to the next-to-leading order (or
$P$-wave) contributions of the conformal spin expansion; the
explicit expressions are given in the appendix; for the details, one
can consult Ref.\cite{BFMS}.

Taking into account the three valence  quark light-cone distribution
amplitudes up to twist-6 and performing the integration over the $x$
in the coordinate space, finally we  obtain the following results,
\begin{eqnarray}
&&z^\mu \Pi_\mu(P,q)\nonumber\\
 &=&  2P\cdot z  N(P)\left\{\int_0^1dt_3 \int_0^{1-t_3}dt_1
 \frac{M_nV_3t_3+m_sV_1}{(q+t_3P)^2}  \right.\nonumber\\
&&+ M_n\int_0^1 d\lambda \int_1^\lambda dt_3 \int_0^{1-t_3}dt_1
\frac{2q\cdot P \lambda
(V_1-V_2-V_3)+\lambda^2M_n^2(V_1-V_2-2V_3+V_4)}{(q+\lambda P)^4}
 \nonumber \\
&&\left.+ M_n^2m_s\int_0^1 d\lambda \lambda\int_1^\lambda dt_3
\int_0^{1-t_3}dt_1
\frac{ -2V_1+V_2+V_3+V_4+V_5}{(q+\lambda P)^4} \right\}\nonumber \\
&+&  2P\cdot z \!\not\! q N(P)\left\{\int_0^1dt_3 \int_0^{1-t_3}dt_1
 \left[\frac{V_1}{(q+t_3P)^2} +\frac{M_n^2V_1^M}{(q+t_3P)^4}\right] \right.\nonumber\\
&&\left.- M_n\int_0^1 d\lambda \int_1^\lambda dt_3
\int_0^{1-t_3}dt_1 \frac{M_n \lambda
(V_1-V_2-V_4)+m_s(V_1-V_2-V_3)}{(q+\lambda P)^4} \right\}
 \nonumber \\
 &+&  2P\cdot z \gamma_5 N(P)\left\{\int_0^1dt_3 \int_0^{1-t_3}dt_1
 \frac{M_nV_3t_3-m_sV_1}{(q+t_3P)^2}  \right.\nonumber\\
&&+ M_n\int_0^1 d\lambda \int_1^\lambda dt_3 \int_0^{1-t_3}dt_1
\frac{2q\cdot P \lambda
(V_1-V_2-V_3)+\lambda^2M_n^2(V_1-V_2-2V_3+V_4)}{(q+\lambda P)^4}
 \nonumber \\
&&\left.- M_n^2m_s\int_0^1 d\lambda \lambda\int_1^\lambda dt_3
\int_0^{1-t_3}dt_1
\frac{ -2V_1+V_2+V_3+V_4+V_5}{(q+\lambda P)^4} \right\}\nonumber \\
&-&  2P\cdot z \!\not\! q \gamma_5 N(P)\left\{\int_0^1dt_3
\int_0^{1-t_3}dt_1
 \left[\frac{V_1}{(q+t_3P)^2} +\frac{M_n^2V_1^M}{(q+t_3P)^4}\right] \right.\nonumber\\
&&\left.- M_n\int_0^1 d\lambda \int_1^\lambda dt_3
\int_0^{1-t_3}dt_1 \frac{M_n\lambda
(V_1-V_2-V_4)-m_s(V_1-V_2-V_3)}{(q+\lambda P)^4} \right\}
 \nonumber \\
&+&\cdots ,
\end{eqnarray}
here the $V_i=V_i(t_1,t_2,t_3)$.

 According to the basic assumption of current-hadron duality in
the QCD sum rules approach \cite{SVZ79}, we insert  a complete
series of intermediate states satisfying the unitarity   principle
with the same quantum numbers as the current operator $\eta(0)$
 into the correlation function in
Eq.(1)  to obtain the hadronic representation. After isolating the
pole term of the lowest  $\Sigma$ state, we obtain the following
result,
\begin{eqnarray}
z^\mu\Pi_\mu(P,q)&=&\frac{\lambda_\Sigma N(P')\langle
N(P')|\bar{s}(0)\!\not\!{z}(1- \gamma_5)
u(0)|N(P)\rangle}{M_\Sigma^2-(q+P)^2}+\cdots
\nonumber \\
&=&2P\cdot z \lambda_\Sigma \frac{ f_1 +g_1\gamma_5}{
M_\Sigma^2-(q+P)^2}N(P)+ \nonumber\\
&& 2P\cdot z\frac{\lambda_\Sigma}{M_n+M_\Sigma}  \frac{
f_2\!\not\!{q} +g_2\!\not\!{q}\gamma_5}{
M_\Sigma^2-(q+P)^2}N(P)+\cdots \, .
\end{eqnarray}
here we have used the definition,
\begin{eqnarray}
&& \langle N(P')|\bar{s}(0)\!\not\!{z}(1- \gamma_5) u(0)|N(P)\rangle
\nonumber\\
&=&N(P')\left\{\!\not\!{z}f_1-i\frac{z_\mu \sigma^{\mu\nu}q_\nu
}{M_\Sigma+M_n}f_2+\frac{q\cdot z }{M_\Sigma+M_n}f_3 \right\}N(P)+
\nonumber \\
&&N(P')\left\{\!\not\!{z}g_1-i\frac{z_\mu \sigma^{\mu\nu}q_\nu
}{M_\Sigma+M_n}g_2+\frac{q\cdot z }{M_\Sigma+M_n}g_3
\right\}\gamma_5N(P) \, .
\end{eqnarray}
Here we choose the light-cone four vector $z_\mu$ with $q\cdot z=0$
and $z^2=0$.  The tensor structures $2 P \cdot z $, $2 P \cdot z
  \gamma_5 $,  $2 P \cdot z \!\not\!{q}  $ and $2 P
\cdot z \!\not\!{q} \gamma_5 $   are chosen to analyze the four
form-factors $f_1(Q^2)$, $f_2(Q^2)$, $g_1(Q^2)$ and $g_2(Q^2)$,
respectively.

The Borel transformation and the continuum states subtraction can be
performed by using the following substitution rules,
\begin{eqnarray}
\int dx \frac{\rho(x)}{(q+xP)^2}&=&-\int_0^1 \frac{dx}{x}
\frac{\rho(x)}{s-{P'}^2}\Rightarrow -\int_{x_0}^1 \frac{dx}{x}
\rho(x)e^{-\frac{s}{M_B^2}} , \nonumber \\
\int dx \frac{\rho(x)}{(q+xP)^4}&=&\int_0^1 \frac{dx}{x^2}
\frac{\rho(x)}{(s-{P'}^2)^2}\Rightarrow  \frac{1}{M_B^2}\int_{x_0}^1
\frac{dx}{x^2}
\rho(x)e^{-\frac{s}{M_B^2}}+\frac{\rho(x_0)e^{-\frac{s_0}{M_B^2}}}{Q^2+x_0^2
M_n^2}\, , \nonumber\\
s&=&(1-x)M_n^2+\frac{(1-x)}{x}Q^2, \nonumber\\
x_0&=&\frac{\sqrt{(Q^2+s_0-M_n^2)^2+4M_n^2Q^2}-(Q^2+s_0-M_n^2)}{2M_n^2}.
\end{eqnarray}
Matching the hadronic  representations and the corresponding
representations at the level of the quark-gluons degrees of freedom
 below the threshold $s_0$,   we obtain the sum rules
for the four form-factors $f_1(Q^2)$, $f_2(Q^2)$, $g_1(Q^2)$ and
$g_2(Q^2)$ ,
\begin{eqnarray}
&&f_1(Q^2)\lambda_\Sigma e^{-\frac{M_\Sigma^2}{M_B^2}}\nonumber\\
&=& - \int_{x_0}^1dt_3\int_0^{1-t_3}dt_1 \exp \left\{-\frac{t_3(1-t_3)M_n^2+(1-t_3)Q^2}{t_3M_B^2}\right\}\left[M_nV_3+\frac{m_sV_1}{t_3}  \right]\nonumber\\
 &+&x_0M_n^2\int_{x_0}^1dt_3\int_0^{1-t_3}dt_1 \exp
 \left\{-\frac{s_0}{M_B^2}\right\} \nonumber \\
 &&\frac{ x_0M_n(V_1-V_2-V_4)-m_s(-2V_1+V_2+V_3+V_4+V_5)  }{x_0^2M_n^2+Q^2}\nonumber\\
&-& \frac{M_n^2}{M_B^2} \int_{x_0}^1d\lambda \int_1^\lambda dt_2\int_0^{1-t_2}dt_1 \frac{1}{\lambda}\exp \left\{-\frac{\lambda(1-\lambda)M^2+(1-\lambda)Q^2}{\lambda M_B^2}\right\}\nonumber \\
&&\left[M_n\lambda(V_1-V_2-V_4)-m_s(-2V_1+V_2+V_3+V_4+V_5)\right]\nonumber\\
&+& M_n \int_{x_0}^1d\lambda \int_1^\lambda dt_2\int_0^{1-t_2}dt_1
(V_1-V_2-V_3) \lambda \nonumber \\
&&\frac{d}{d\lambda} \frac{1}{\lambda}\exp
\left\{-\frac{\lambda(1-\lambda)M^2+(1-\lambda)Q^2}{\lambda
M_B^2}\right\}\, ;
\end{eqnarray}

\begin{eqnarray}
&&g_1(Q^2)\lambda_\Sigma e^{-\frac{M_\Sigma^2}{M_B^2}}\nonumber\\
&=& - \int_{x_0}^1dt_3\int_0^{1-t_3}dt_1 \exp \left\{-\frac{t_3(1-t_3)M_n^2+(1-t_3)Q^2}{t_3M_B^2}\right\}\left[M_nV_3-\frac{m_sV_1}{t_3}  \right]\nonumber\\
 &+&x_0M_n^2\int_{x_0}^1dt_3\int_0^{1-t_3}dt_1 \exp
 \left\{-\frac{s_0}{M_B^2}\right\} \nonumber \\
 &&\frac{ x_0M_n(V_1-V_2-V_4)+m_s(-2V_1+V_2+V_3+V_4+V_5)  }{x_0^2M_n^2+Q^2}\nonumber\\
&-& \frac{M_n^2}{M_B^2} \int_{x_0}^1d\lambda \int_1^\lambda dt_2\int_0^{1-t_2}dt_1 \frac{1}{\lambda} \exp \left\{-\frac{\lambda(1-\lambda)M^2+(1-\lambda)Q^2}{\lambda M_B^2}\right\}\nonumber \\
&&\left[\lambda M_n(V_1-V_2-V_4)+m_s(-2V_1+V_2+V_3+V_4+V_5)\right]\nonumber\\
&+& M_n \int_{x_0}^1d\lambda \int_1^\lambda dt_2\int_0^{1-t_2}dt_1
(V_1-V_2-V_3) \lambda \nonumber\\
&&\frac{d}{d\lambda} \frac{1}{\lambda}\exp
\left\{-\frac{\lambda(1-\lambda)M^2+(1-\lambda)Q^2}{\lambda
M_B^2}\right\}\, ;
\end{eqnarray}

\begin{eqnarray}
&&f_2(Q^2)\frac{\lambda_\Sigma}{M_\Sigma+M_n} e^{-\frac{M_\Sigma^2}{M_B^2}}\nonumber\\
&=& - \int_{x_0}^1dt_3\int_0^{1-t_3}dt_1 \exp
\left\{-\frac{t_3(1-t_3)M_n^2+(1-t_3)Q^2}{t_3M_B^2}\right\}
\left[\frac{V_1}{t_3}-\frac{M_n^2V_1^M}{t_3^2 M_B^2}  \right]\nonumber\\
 &+&M_n\int_{x_0}^1dt_3\int_0^{1-t_3}dt_1 \exp
 \left\{-\frac{s_0}{M_B^2}\right\} \frac{ x_0M_n(V_1-V_2-V_4)+m_s(V_1-V_2-V_3)  }{x_0^2M_n^2+Q^2}\nonumber\\
&-& \frac{M_n}{M_B^2} \int_{x_0}^1d\lambda \int_1^\lambda dt_2\int_0^{1-t_2}dt_1 \exp \left\{-\frac{\lambda(1-\lambda)M^2+(1-\lambda)Q^2}{\lambda M_B^2}\right\}\nonumber \\
&&\frac{\lambda M_n(V_1-V_2-V_4)+m_s(V_1-V_2-V_3)}{\lambda^2}\nonumber\\
&+&M_n^2\int_0^{1-x_0}dt_1 \frac{V_1^M }{x_0^2M_n^2+Q^2}\exp
 \left\{-\frac{s_0}{M_B^2}\right\} \,;
\end{eqnarray}

\begin{eqnarray}
&&g_2(Q^2)\frac{\lambda_\Sigma}{M_\Sigma+M_n} e^{-\frac{M_\Sigma^2}{M_B^2}}\nonumber\\
&=&  \int_{x_0}^1dt_3\int_0^{1-t_3}dt_1 \exp
\left\{-\frac{t_3(1-t_3)M_n^2+(1-t_3)Q^2}{t_3M_B^2}\right\}\left[\frac{V_1}{t_3}-\frac{M_n^2V_1^M}{t_3^2 M_B^2}  \right]\nonumber\\
 &-&M_n\int_{x_0}^1dt_3\int_0^{1-t_3}dt_1 \exp
 \left\{-\frac{s_0}{M_B^2}\right\} \frac{ x_0M_n(V_1-V_2-V_4)-m_s(V_1-V_2-V_3)  }{x_0^2M_n^2+Q^2}\nonumber\\
&+& \frac{M_n}{M_B^2} \int_{x_0}^1d\lambda \int_1^\lambda dt_2\int_0^{1-t_2}dt_1 \exp \left\{-\frac{\lambda(1-\lambda)M^2+(1-\lambda)Q^2}{\lambda M_B^2}\right\}\nonumber \\
&&\frac{\lambda M_n(V_1-V_2-V_4)-m_s(V_1-V_2-V_3)}{\lambda^2}\nonumber\\
&-&M_n^2\int_0^{1-x_0}dt_1 \frac{V_1^M }{x_0^2M_n^2+Q^2}\exp
 \left\{-\frac{s_0}{M_B^2}\right\} \, .
\end{eqnarray}

 In the chiral limit $m_s\rightarrow 0$, the $f_1(Q^2)=g_1(Q^2)$ and
 $f_2(Q^2)=-g_2(Q^2)$.

\section{Numerical results and discussions}
The input parameters have to be specified  before the numerical
analysis. We choose the suitable range  for the Borel parameter
$M_B$, $2.0GeV^2<M_B^2<3.0GeV^2$. In this range, the Borel parameter
$M_B$ is small enough  to warrant the higher mass resonances and
 continuum states are  suppressed sufficiently, on the other hand,  it is
  large enough to warrant the convergence of the
light-cone expansion with increasing twists in the perturbative QCD
calculation \cite{Ioffe}. The numerical results indicate that in
this range the  four form-factors $f_1(Q^2)$, $f_2(Q^2)$, $g_1(Q^2)$
and $g_2(Q^2)$ are almost independent on the Borel parameter $M_B$,
in this article, we choose the special value $M_B^2=2.5GeV^2$ for
simplicity.

We choose the standard  value for the  threshold parameter $s_0$,
$s_0=3.2GeV^2$,  to
 subtract the contributions from the higher resonances and
 continuum states \cite{s0};  it is large enough to
take into account  all contributions from the $\Sigma$ baryon.   For
$Q^2>3GeV^2$, $x\geq x_0\geq0.5$,  with the intermediate and large
space-like momentum $Q^2$, the end-point (soft) contributions (or
the Feynman mechanism) are dominant, it is consistent with the
growing consensus that the onset of the perturbative QCD region in
exclusive processes is postponed to very large energy scales. We
perform the operator product expansion at the regions $Q^2\gg 0$ and
$(q+P)\ll 0$, and obtain the sum rules in Eqs.(11-14),  the
form-factors $f_1(Q^2)$, $g_1(Q^2)$, $f_2(Q^2)$ and $g_2(Q^2)$ make
sense at the regions, for example,  $Q^2> 3GeV^2$, with low momentum
transfers, the operator product expansion is questionable. We
extrapolate the values of the $Q^2$ to zero, the functions
$f_1(Q^2)$, $g_1(Q^2)$, $f_2(Q^2)$ and $g_2(Q^2)$ happen have rather
good behavior at lower momentum transfers \footnote{We can borrow
some ideas from the electromagnetic form-factor of the $\pi$-photon
$f_{\gamma^* \pi^0 }(Q^2)$, the  value of the $f_{\gamma^* \pi^0
}(0)$ is fixed by the partial conservation of the axial current and
the effective  anomaly lagrangian, $f_{\gamma^* \pi^0 }(0) =
\frac1{\pi f_{\pi}}$, in the limit large-$Q^2$, the perturbative QCD
predicts that $f_{\gamma^* \pi^0 }(Q^2) =4\pi f_{\pi}/Q^2 $.
 The Brodsky-Lepage interpolation formula \cite{BJ81}
 \begin{eqnarray}
f_{\gamma^* \pi^0 }(Q^2) = \frac{1}{ \pi f_{\pi} \left [1+Q^2/(4
\pi^2 f_{\pi}^2) \right ]} =\frac{1}{ \pi f_\pi (1+Q^2/s_0) }
\nonumber
\end{eqnarray}
 can reproduce both the value of $Q^2 =0$
  and the  behavior of large-$Q^2$, the energy scale $s_0$ ($s_0 = 4 \pi^2
f_{\pi}^2 \approx 0.67 \, GeV^2$) is numerically   close to   the
squared mass of the $\rho$ meson, $m_{\rho}^2 \approx 0.6 \, GeV^2$.
The Brodsky-Lepage interpolation formula is similar to the result of
the vector meson dominance, $f_{\gamma^* \pi^0 }(Q^2) = 1/\left\{\pi
f_\pi (1+Q^2/m_{\rho}^2)\right\}$. In the vector meson dominance
approach, the calculation is performed at the time-like energy scale
$q^2<1GeV^2$ and the electromagnetic  current is saturated by the
vector meson $\rho$, where the mass $m_{\rho}$ serves as a parameter
determining the pion charge radius. With a slight modification of
the mass parameter, $m_\rho=\Lambda_\pi=776MeV$, the experimental
data can be well described by the single-pole formula  at the
interval $Q^2=(0-10)GeV^2$ \cite{CLEO97}. In this article, the four
form-factors have satisfactory behaviors  at large $Q^2$  which are
expected by the naive power counting rules and have finite values at
$Q^2=0$, the analytical expressions $f_1(Q^2)$, $f_2(Q^2)$,
$g_1(Q^2)$ and $g_2(Q^2)$ can be taken as some Brodsky-Lepage  type
interpolation formulaes, although they are calculated at rather
large $Q^2$, the extrapolation to the lower energy transfers has no
solid theoretical foundation.}.

The mass of the $s$ quark is chosen to be $m_s=140MeV$ at the energy
scale $\mu=1GeV$. In the absence of second class currents
\cite{Weinberg58} the form-factor $g_2(Q^2)$  vanishes in the
$SU(3)$ symmetry limit. The neutral currents $A^3_\mu$ and $A^8_\mu$
belong to the same octet as the weak axial currents  are even under
charge conjugation, their matrix elements cannot contain a
weak-electricity term, which is $C$-odd. The vanishing of the weak
electricity in the proton and neutron matrix elements of the
$A^3_\mu,\, A^8_\mu$ implies the vanishing of the $g_2(0)$ in the
$SU(3)$ symmetry limit. As the current masses of the $u$ and $d$
quarks are very small, the $SU(3)$ symmetry breaking effects can be
taken into account by the non-vanishing $m_s$. In calculation, we
observe that central value $g_2^{\chi}(0)=2.31$ in the chiral limit,
the inclusion of the terms proportional to the $m_s$ can not change
the result drastically,  $g_2(0)=1.92$.

 The parameters in the light-cone distribution amplitudes  $ \phi_3^0 $,
  $\phi_6^0$, $\phi_4^0$, $\phi_5^0$,  $\xi_4^0$, $\xi_5^0$, $\psi_4^0$,
$\psi_5^0 $,  $ \phi_3^-$, $\phi_3^+$, $\phi_4^-$, $\phi_4^+$,
$\psi_4^-$, $\psi_4^+$, $\xi_4^-$, $\xi_4^+$, $\phi_5^-$,
$\phi_5^+$, $\psi_5^-$, $\psi_5^+$, $\xi_5^-$ ,$ \xi_5^+$,
$\phi_6^-$, $\phi_6^+ $ are scale dependent and can be calculated
with the corresponding QCD sum rules. They are functions of eight
independent parameters $f_N$, $\lambda_1$, $\lambda_2$, $V_1^d$,
$A_1^u$, $f^d_1$, $f^d_2$ and $f^u_1$,  the three parameters $f_N$,
$\lambda_1$ and $\lambda_2$ are related to the leading order (or
$S$-wave) contributions in the conformal spin expansion,  the
remaining five parameters $V_1^d$, $A_1^u$, $f^d_1$, $f^d_2$ and
$f^u_1$ are related to the next-to-leading order (or $P$-wave)
contributions in the conformal spin expansion; the explicit
expressions are presented in the appendix, for detailed and
systematic studies about this subject, one can consult
Ref.\cite{BFMS}.  Here we take the values at the energy scale
$\mu=1GeV$ and neglect the  evolution with the energy scale $\mu$
for simplicity, the values of the eight independent parameters are
taken as $f_N=(5.0\pm 0.5)\times 10^{-3} GeV^2$, $\lambda_1=-(2.7\pm
0.9)\times 10^{-2}GeV^2$, $\lambda_2=(5.1\pm 1.9)\times
10^{-2}GeV^2$, $V_1^d=0.23\pm 0.03$, $A_1^u=0.38\pm 0.15$
\cite{BFMS}, $f_1^d=0.40\pm 0.05$, $f_2^d=0.22\pm 0.05$ and
$f_1^u=0.07\pm 0.05  $ \cite{Braun06}. In estimating those
coefficients with the QCD sum rules, only the first few moments are
taken into account, the values are not very accurate. In the limit
$Q^2\rightarrow \infty$, the five parameters related to the
light-cone distribution amplitudes with the $P$-wave conformal spin
take the asymptotic values $f^d_1=\frac{3}{10}$,
$f^d_2=\frac{4}{15}$, $f^u_1=\frac{1}{10}$, $A^u_1=0$ and
$V^d_1=\frac{1}{3}$.

In numerical analysis, we observe that  the form-factors $f_1(Q^2)$
and $g_1(Q^2)$ are sensitive to the   parameter $\lambda_1$, small
variations of the  parameter can lead to large changes of the
values, the  form-factors $f_2(Q^2)$ and $g_2(Q^2)$ are sensitive to
the three parameters $f_N$, $\lambda_1$ and $f^d_1$, small
variations of those parameters  can lead to relatively large changes
of the values. The   large uncertainties can impair the predictive
ability of the sum rules,  the parameters $\lambda_1$, $f^d_1$ and
$f_N$ should be refined to make robust predictions, in
Refs.\cite{WangWY} \footnote{In Refs.\cite{WangWY}, we have
neglected some terms which we take it for granted as un-important in
performing the operator product expansion,  the predictive power may
be impaired to  some extent. In this article, we use the chiral
current to study the vector and axial-vector form-factors in an
unified way. Some terms may be canceled out with each other in
performing the operator expansion with the chiral current, and this
approach may result in more reasonable values; it is indeed the case
for the vector form-factor of the mesons, for example, the $B \to
\pi$ form-factor in Ref.\cite{WZH03}. It is interesting to study the
form-factors of the nucleons with the chiral currents. }, we observe
that the scalar form-factor of the nucleon is sensitive to the four
parameters $\lambda_1$, $f^d_1$, $f^d_2$ and $f^u_1$, and the axial
and induced pseudoscalar form-factors are sensitive to the four
parameters $\lambda_1$, $f^d_1$, $f_N$ and $f^u_1$, so refining the
three parameters $\lambda_1$, $f_N$,  and $f^d_1$ is of great
importance; however, it is difficulty to
 pin down the uncertainties. The final numerical values of  the four
form-factors $f_1(Q^2)$, $f_2(Q^2)$, $g_1(Q^2)$ and $g_2(Q^2)$ at
$0<Q^2<5.5GeV^2$ are plotted in the Fig.1.
\begin{figure}
\centering
  \includegraphics[totalheight=7cm,width=7cm]{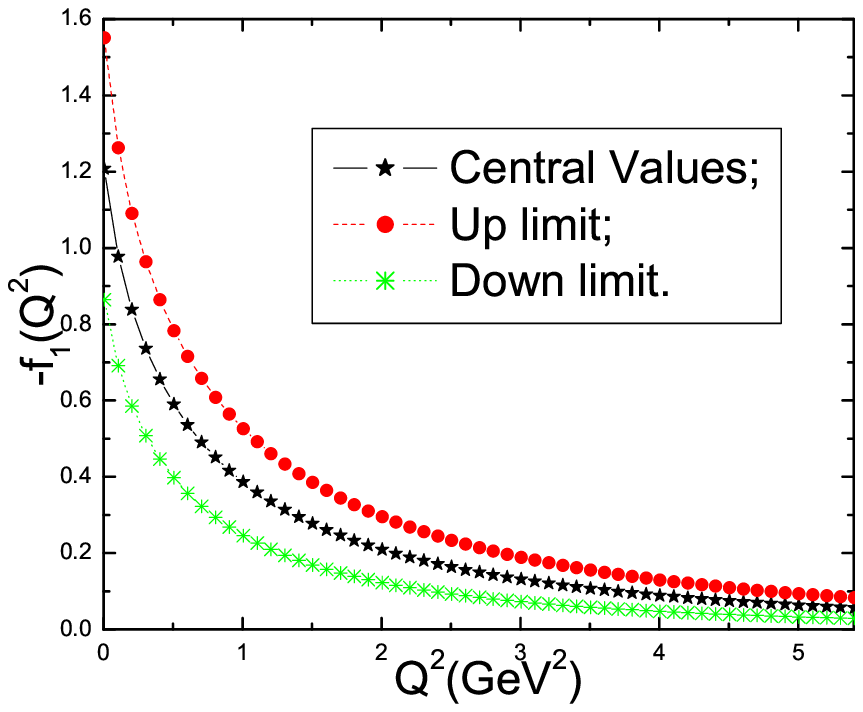}
  \includegraphics[totalheight=7cm,width=7cm]{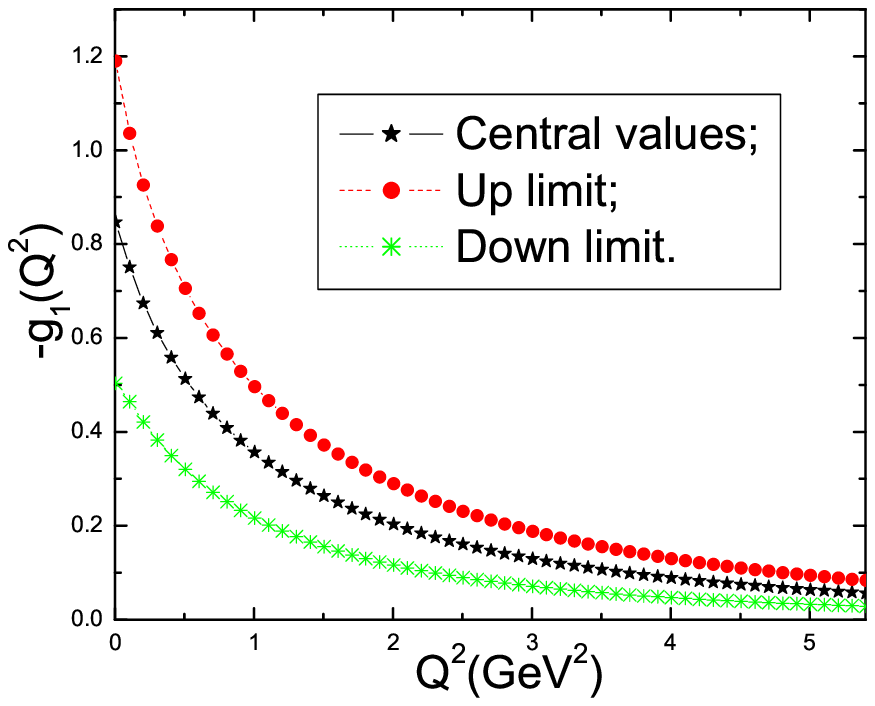}
  \includegraphics[totalheight=7cm,width=7cm]{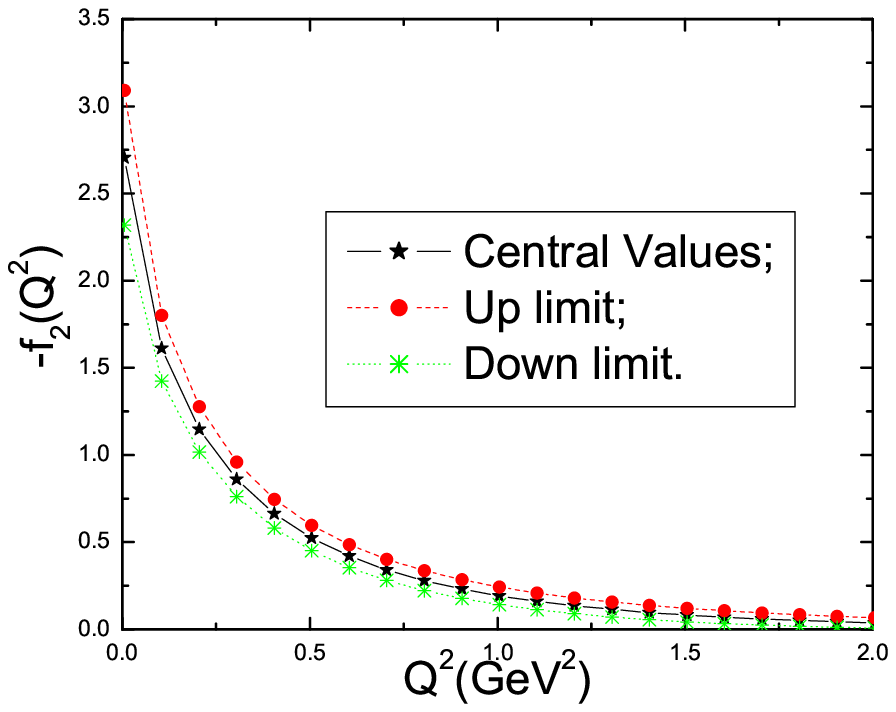}
\includegraphics[totalheight=7cm,width=7cm]{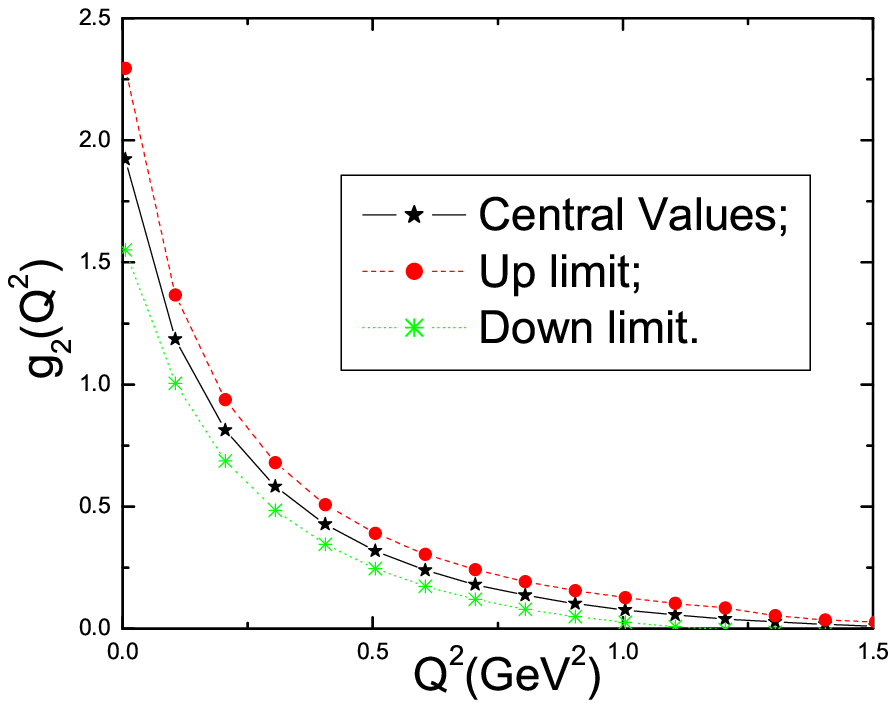}
 \caption{The   $f_1(Q^2)$, $g_1(Q^2)$, $f_2(Q^2)$ and $g_2(Q^2)$ with the parameter $Q^2$. }
\end{figure}
The central values of the four form-factors can be approximately
fitted into the double-pole formula,
\begin{eqnarray}
F(Q^2)=\frac{F(0)}{\left(1+\frac{Q^2}{m_A^2}
\right)\left(1+\frac{Q^2}{m_B^2}\right)} \, ,
\end{eqnarray}
here the $F(Q^2)$ stand for the $f_1(Q^2)$, $f_2(Q^2)$, $g_1(Q^2)$
and $g_2(Q^2)$, the corresponding values of the parameters $m_A$ and
$m_B$ are listed in the Table 1.
\begin{table}
\begin{center}
\begin{tabular}{c|c|c}
\hline\hline
      $F(Q^2)$ &$m_A(GeV)$&$m_B(GeV)$    \\ \hline
      $f_1$& $0.95$&$1.55$  \\      \hline
$f_2$& $0.55$&$0.60$  \\  \hline
$g_1$&$1.10$&$1.90$ \\
\hline $g_2$&$0.43$&$0.75$  \\      \hline
      \hline
\end{tabular}
\end{center}
\caption{ Numerical values of the  parameters $m_A$ and $m_B$ . }
\end{table}
 From the numerical values
\begin{eqnarray}
f_1(0)&=&-(1.21\pm0.34) \, , \nonumber\\
g_1(0)&=&-(0.85\pm0.34) \, , \nonumber\\
g_1(0)/f_1(0)_{c}&=&0.70  \, , \nonumber\\
f_2(0)&=&-(2.70\pm0.39) \, , \nonumber\\
f_2(0)/f_1(0)_{c}&=&2.23 \, ,  \nonumber\\
g_2(0)&=&1.92\pm0.37 \, , \nonumber \\
g_2(0)/f_1(0)_{c}&=&-1.59 \, , \nonumber\\
f_1^{\chi}(0)&=&g_1^{\chi}(0)=-1.03\, , \nonumber \\
f_2^{\chi}(0)&=&-g_2^{\chi}(0)=-2.31\, , \nonumber \\
f_1^{asp}(0)&=&-1.11 \, , \nonumber\\
g_1^{asp}(0)&=&-0.85 \, , \nonumber\\
f_2^{asp}(0)&=&-2.67 \, , \nonumber\\
g_2^{asp}(0)&=&1.20 \, ,
\end{eqnarray}
 we can see that they are compatible with the experimental data and
 theoretical estimations (in magnitude), $f_2(0)/f_1(0)=-1.71
\pm 0.12 \pm 0.23$ (experimental data) \cite{HsuehExp88};
$g_1(0)/f_1(0)=-0.340\pm0.017$ (theoretical
estimation)\cite{Cabibbo03}; $f_1(0)=-0.988\pm0.029\pm0.040$,
$g_1(0)/f_1(0)=-0.287\pm0.052$, $f_2(0)/f_1(0)=-1.52\pm0.81$,
$g_2(0)/f_1(0)=0.63\pm0.26$ (lattice simulation) \cite{Latt06}. Here
 the $c$ , $asp$ and $\chi$ stand for the cental values, the
asymptotic values and the values in the chiral limit, respectively.
The discrepancy may be due to the perturbative $\alpha_s$
corrections, additional valence gluons and quark-antiquark pairs.
The consistent and complete LCSR analysis should take into account
the contributions from the perturbative $\alpha_s$ corrections, the
distribution amplitudes with additional valence gluons and
quark-antiquark pairs, and improve the parameters which enter in the
LCSRs.

\section{Conclusion }

In this work, we  calculate the four form-factors
   $f_1(Q^2)$, $f_2(Q^2)$, $g_1(Q^2)$ and $g_2(Q^2)$ of the $\Sigma \to n$ in the
  framework of the LCSR approach up
 to twist-6 three valence quark light-cone distribution amplitudes.
 The $f_1(0)$ is the basic
input parameter in extracting the CKM matrix element $|V_{us}|$ from
the hyperon decays.
  The  four form-factors $f_1(Q^2)$, $f_2(Q^2)$, $g_1(Q^2)$ and $g_2(Q^2)$ at intermediate and
large momentum transfers with $Q^2> 3 GeV^2$ have significant
contributions from the end-point (soft) terms. The form-factors
$f_1(Q^2)$ and $g_1(Q^2)$ are sensitive to the   parameter
$\lambda_1$, small variations of the parameter can lead to large
changes of the values, the form-factors $f_2(Q^2)$ and $g_2(Q^2)$
are sensitive to the three parameters $f_N$, $\lambda_1$ and
$f^d_1$, small variations of those parameters  can lead to
relatively large changes of the values. The   large uncertainties
can impair the predictive ability of the sum rules,  the parameters
$\lambda_1$, $f_N$ and $f^d_1$ should be refined to make robust
predictions. The numerical values of the four form-factors
   $f_1(0)$, $f_2(0)$, $g_1(0)$ and $g_2(0)$  are
 compatible with the experimental data and  theoretical
 calculations (in magnitude).
 The consistent and complete LCSR analysis should take into account
the contributions from the perturbative $\alpha_s$ corrections, the
distribution amplitudes with additional valence gluons and
quark-antiquark pairs, and improve the parameters which enter in the
LCSRs.

\section*{Acknowledgment}
This  work is supported by National Natural Science Foundation,
Grant Number 10405009,  and Key Program Foundation of NCEPU.
\appendix
\section*{Appendix}
 \begin{eqnarray}
V_1(x_i,\mu)&=&120x_1x_2x_3[\phi_3^0(\mu)+\phi_3^+(\mu)(1-3x_3)],\nonumber\\
V_2(x_i,\mu)&=&24x_1x_2[\phi_4^0(\mu)+\phi_3^+(\mu)(1-5x_3)],\nonumber\\
V_3(x_i,\mu)&=&12x_3\{\psi_4^0(\mu)(1-x_3)+\psi_4^-(\mu)[x_1^2+x_2^2-x_3(1-x_3)]
\nonumber\\&&+\psi_4^+(\mu)(1-x_3-10x_1x_2)\},\nonumber\\
V_4(x_i,\mu)&=&3\{\psi_5^0(\mu)(1-x_3)+\psi_5^-(\mu)[2x_1x_2-x_3(1-x_3)]
\nonumber\\&&+\psi_5^+(\mu)[1-x_3-2(x_1^2+x_2^2)]\},\nonumber\\
V_5(x_i,\mu)&=&6x_3[\phi_5^0(\mu)+\phi_5^+(\mu)(1-2x_3)],\nonumber\\
V_6(x_i,\mu)&=&2[\phi_6^0(\mu)+\phi_6^+(\mu)(1-3x_3)].\nonumber
\end{eqnarray}

\begin{eqnarray}
 \mathcal{V}_1^{u}(x_3) &=&\frac{x_3^2}{24}(\lambda_1
C_\lambda^u+f_N C_f^u),\nonumber\\
C_\lambda^u&=& -(1-x_3)[11+131\,x_3-169x_3^2+63x_3^3-30\,f_1^d
\,(3+11x_3-17x_3^2+7x_3^3)]\nonumber \\
&&-12\,(3-10\,f_1^d)\,\ln x_3,\nonumber\\
C_f^u&=& -( 1 - x_3 )
\,[1441+505x_3-3371x_3^2+3405x_3^3-1104x_3^4-24V_1^d\nonumber\\
&&(207-3x_3-368x_3^2+412x_3^3-138x_3^4)]  - 12(73-220\,V_1^d)\,\ln
x_3,\nonumber
\end{eqnarray}

\begin{eqnarray}
\phi_3^0 = \phi_6^0 = f_N \,,\hspace{0.3cm} &\qquad& \phi_4^0 =
\phi_5^0 = \frac{1}{2} \left(\lambda_1 + f_N\right) \,,
\nonumber \\
\xi_4^0 = \xi_5^0 = \frac{1}{6} \lambda_2\,, &\qquad& \psi_4^0  =
\psi_5^0 = \frac{1}{2}\left(f_N - \lambda_1 \right)  \,. \nonumber
\end{eqnarray}
\begin{eqnarray}
\tilde\phi_3^- &=& \frac{21}{2} A_1^u,\nonumber\\
\tilde\phi_3^+ &=& \frac{7}{2} (1 - 3 V_1^d),\nonumber\\
\phi_4^- &=& \frac{5}{4} \left(\lambda_1(1- 2 f_1^d -4 f_1^u) + f_N(
2 A_1^u - 1)\right) \,,
\nonumber \\
\phi_4^+ &=& \frac{1}{4} \left( \lambda_1(3- 10 f_1^d) - f_N( 10
V_1^d - 3)\right)\,,
\nonumber \\
\psi_4^- &=& - \frac{5}{4} \left(\lambda_1(2- 7 f_1^d + f_1^u) +
f_N(A_1^u + 3 V_1^d - 2)\right) \,,
\nonumber \\
\psi_4^+ &=& - \frac{1}{4} \left(\lambda_1 (- 2 + 5 f_1^d + 5 f_1^u)
+ f_N( 2 + 5 A_1^u - 5 V_1^d)\right)\,,
\nonumber \\
\xi_4^- &=& \frac{5}{16} \lambda_2(4- 15 f_2^d)\,,
\nonumber \\
\xi_4^+ &=& \frac{1}{16} \lambda_2 (4- 15 f_2^d)\,,\nonumber
\end{eqnarray}
\begin{eqnarray}
\phi_5^- &=& \frac{5}{3} \left(\lambda_1(f_1^d - f_1^u) + f_N( 2
A_1^u - 1)\right) \,,
\nonumber \\
\phi_5^+ &=& - \frac{5}{6} \left(\lambda_1 (4 f_1^d - 1) + f_N( 3 +
4 V_1^d)\right)\,,
\nonumber \\
\psi_5^- &=& \frac{5}{3} \left(\lambda_1 (f_1^d - f_1^u) + f_N( 2 -
A_1^u - 3 V_1^d)\right)\,,
\nonumber \\
\psi_5^+ &=& -\frac{5}{6} \left(\lambda_1 (- 1 + 2 f_1^d +2 f_1^u) +
f_N( 5 + 2 A_1^u -2 V_1^d)\right)\,,
\nonumber \\
\xi_5^- &=& - \frac{5}{4} \lambda_2 f_2^d\,,
\nonumber \\
\xi_5^+ &=&  \frac{5}{36} \lambda_2 (2 - 9 f_2^d)\,,
\nonumber \\
\phi_6^- &=& \phantom{-}\frac{1}{2} \left(\lambda_1 (1- 4 f_1^d - 2
f_1^u) + f_N(1 +  4 A_1^u )\right) \,,
\nonumber \\
\phi_6^+ &=& - \frac{1}{2}\left(\lambda_1  (1 - 2 f_1^d) + f_N ( 4
V_1^d - 1)\right)\,. \nonumber
\end{eqnarray}

\end{document}